\documentclass[12pt,preprint]{aastex}

\slugcomment{\today}

\begin{document}

\title{A New Brown Dwarf Desert?\\
A Scarcity of Wide Ultracool Binaries}

\author{Peter R.\ Allen}
\affil{\small Pennsylvania State University, 525 Davey Lab, University Park PA 16802; pallen@astro.psu.edu}
\author{David W.\ Koerner}
\affil{\small Northern Arizona University, Dept.\ of Physics and Astronomy, PO Box 6010, Flagstaff, AZ 86011-6010; koerner@physics.nau.edu}
\author{Michael W.\ McElwain}
\affil{\small Department of Physics and Astronomy, UCLA, Los Angeles, CA 90095-1592; mcelwain@astro.ucla.edu}
\author{Kelle L.\ Cruz}
\affil{\small American Museum of Natural History, Dept.\ of Astrophysics, Central Park West at 79th Street, New York, New York, 10024-5192; kelle@amnh.org}
\author{I.\ Neill Reid}
\affil{\small Space Telescope Science Institute, 3700 San Martin Drive, Baltimore, MD 21218; inr@stsci.edu}

\begin{abstract}

We present the results of a deep-imaging search for wide companions to low-mass stars and brown dwarfs using NSFCam on IRTF.  We searched a sample of 132 M7-L8 dwarfs to magnitude limits of $J \sim 20.5$ and $K \sim 18.5$, corresponding to secondary-primary mass ratios of $\sim 0.5$.  No companions were found with separations between $2{\arcsec}$ to $31{\arcsec}$ ($\sim$40~AU to $\sim$1000~AU).  This null result implies a wide companion frequency below $2.3\%$ at the $95\%$ confidence level within the sensitivity limits of the survey.  Preliminary modeling efforts indicate that we could have detected $85\%$ of companions more massive than $0.05~M_{\odot}$ and $50\%$ above $0.03~M_{\odot}$.

\end{abstract}


\section{Introduction}

Low-mass stars and brown dwarfs are likely to be the most numerous constituents of the solar neighborhood.  Over the last few years, several major surveys for these objects have been undertaken, aimed at measuring their numbers and discovering their origins.  More than 350 L dwarfs and as many late-type M dwarfs have been discovered as a result of these projects \citep{xd99,kirk00,xf00,kc03,pb03}.  The current consensus is that the increasing number of objects per unit mass seen in high and intermediate mass stars begins to flatten noticeably near the substellar limit, $0.08~M_{\odot}$ \citep{kru02,burg04,a05}. 

The origin of these low-mass ultracool dwarfs remains in question.  The standard scenario envisions brown dwarfs forming in isolation, like higher mass stars, with the lower mass of the final product reflecting the smaller reservoir of material.  However, a recent suggestion is that ultracool dwarfs have low masses because they are ejected from small stellar groups \citep{rc01}, rather than forming in isolation, as theorized for higher mass stars.  This removes low-mass pre-stellar cores from the star forming cloud, truncating the accretion process and leading to the formation of very low-mass stars or brown dwarfs.

The frequency of ultracool binary systems and the distribution of their properties (mass ratios, separations, orbital eccentricities, etc.) provide constraints on formation models. The ejection scenario, for example, predicts a low binary frequency and few, if any, wide systems.  High spatial resolution observations, with both the Hubble Space Telescope \citep{mar98,inr01,gizis03,burg03,bouy03} and ground-based high-resolution cameras and adaptive optics systems \citep{k99,close03,sig03}, have shown that $\sim$20\% of ultracool dwarfs are binary systems.  None of these binaries has a separation that exceeds 15~AU.  Only one field ultracool dwarf system has been discovered to date with a separation greater than 15~AU \citep{bill05}.  This is in contrast to ultracool dwarfs with higher mass primaries: VB~10, for example, the archetypal late-type M dwarf, lies 400~AU from its primary, the M3 dwarf, Gl~752A \citep{vb}, while the nearby T dwarfs, Gl~229B, Gl~570D and $\epsilon$ Indi Bab, are all wide components in multiple systems.  A handful of wide, {\it young} ultracool binaries have been discovered.  \citet{lu04} has identified a pair of late-type M dwarfs separated by 240 AU in $\rho$ Ophiuchus, and \citet{chau} have discovered a very low-mass brown dwarf companion of the TW~Hya member 2M1207, with a separation of 60~AU.

This paper describes our survey to determine if there are any wide ultracool dwarf binary systems in the field.  We obtained deep, multi-epoch $J$ and $K$ images of 132 isolated dwarfs with spectral types from M7 to L8 to an absolute $J$-band magnitude of $\sim$17.5.  We searched for candidate companions using photometric criteria to verify the nature of those candidates.  Section 2 details the target selection, the imaging observations, the reduction and analysis of the imaging data, section 3 describes the candidate selection process and follow-up observations, and section 4 summarizes our results and discusses their implications.

\section{Target Selection, Observations, and Data Reduction}

\subsection{Target Selection and Sample Information}

Our sample is a subset of the first ultracool dwarf surveys \citep{kp97,kp99,kirk00,newn,xd99,xf00}.  Those initial surveys tended to concentrate on brighter candidates, particularly in follow-up observations of extremely red DENIS and 2MASS sources. As a result, our sample is effectively magnitude-limited and is therefore likely to include a higher proportion of unresolved close binary systems than a volume-limited sample \citep{burg03}. This bias is not directly relevant to the prime purpose of the present survey, which aims to determine the frequency of wide companions to ultracool dwarfs.  

\begin{figure}
\centering
\rotatebox{-90}{
\epsscale{0.8}
\plotone{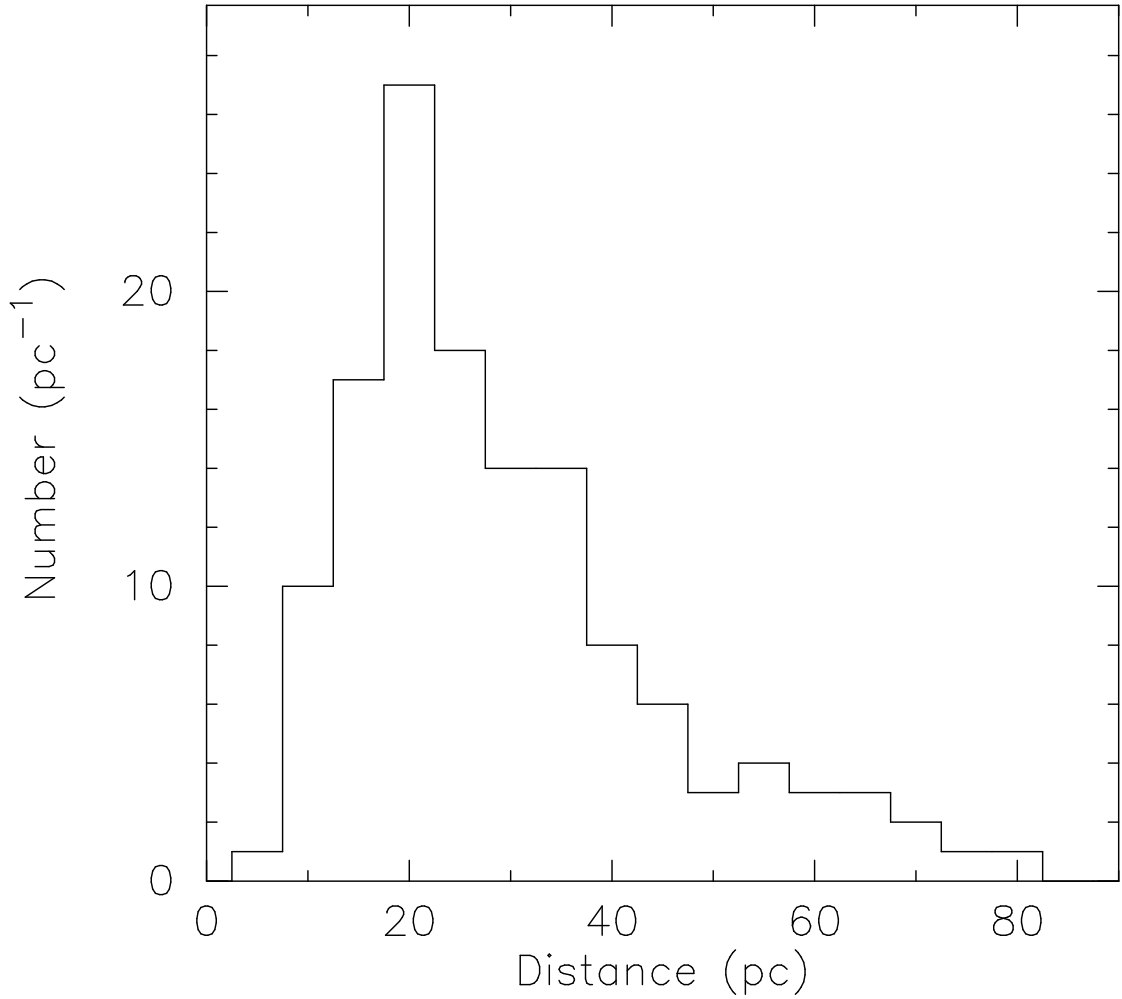}}
\caption{Histogram of the distance estimates for all 132 target primaries in our IRTF sample.  Estimates were obtained through a combination of trigonometric parallaxes \citep{dahn} and calibrated spectrophotometric relations \citep{kc03}.}
\label{fig:dislim}
\end{figure}

\begin{figure}
\centering
\rotatebox{-90}{
\epsscale{0.8}
\plotone{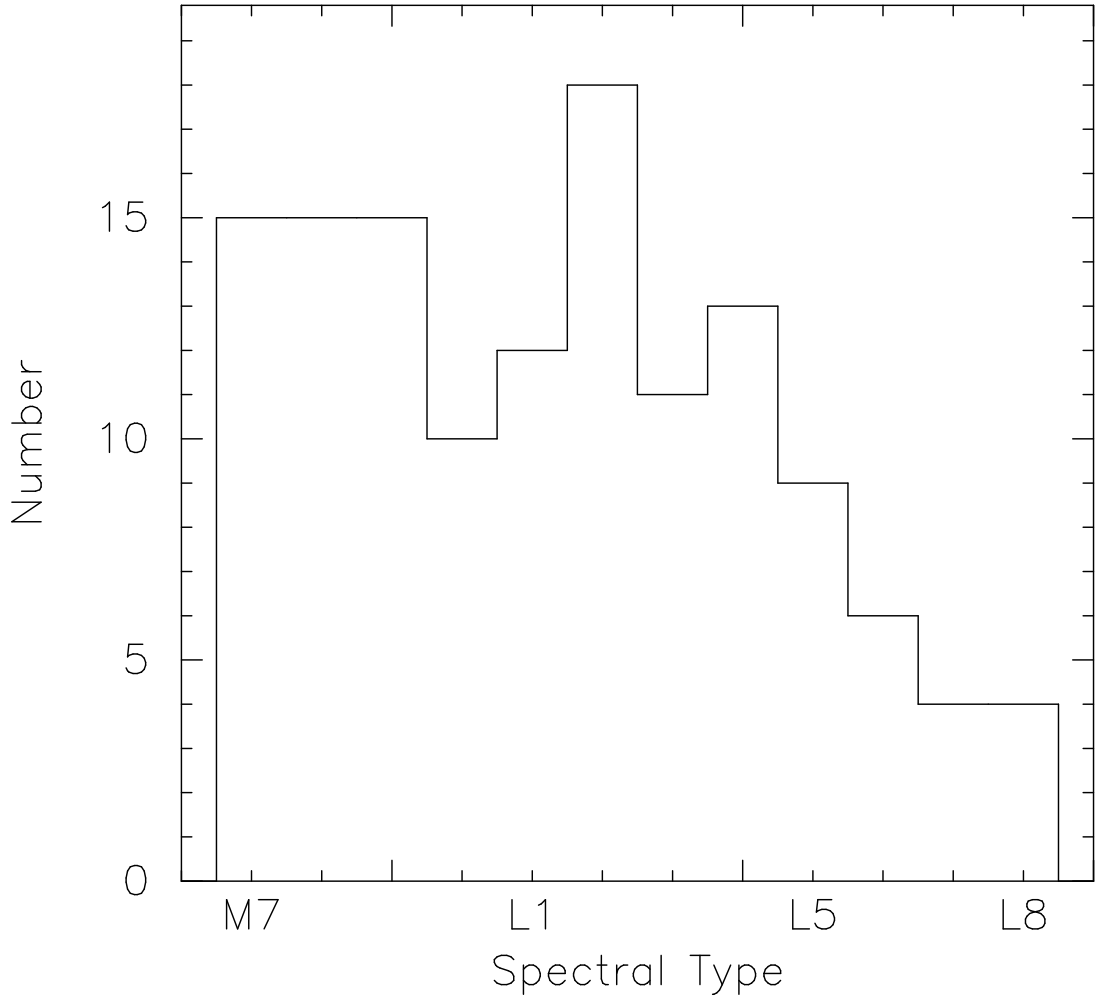}}
\caption{Histogram of the spectral types for all 132 target primaries in our IRTF sample.  Spectral types were obtained from initial discovery papers (see Table \ref{tab:obs}).}
\label{fig:sptdis}
\end{figure}

Figure \ref{fig:dislim} displays the distance estimates of the 132 targets observed in the present program.  Those distances are based primarily on the spectroscopic parallaxes of \citet{kc03}, although a few objects have trigonometric parallax measurements (see Section \ref{sec:selcrit}).  Most candidates are within 30~pc of the Sun.  Figure \ref{fig:sptdis} shows the spectral type distribution, which is essentially flat from late-M to mid-L.  The drop in numbers at later types reflects the relatively small numbers of those objects in the initial surveys.  Thus, while the target sample is not statistically complete, it is representative of the nearby ultracool dwarf population.

\subsection{IRTF Data}
\label{sec:datared}

We observed all 132 targets using NSFCam on NASA's Infrared Telescope Facility (IRTF) \citep{nsfcam}.  The initial observations were obtained over four epochs, August 2000, May 2001, October 2001, and February 2002.  We imaged each target at least once in both the $J$ and $K$ bands (Table \ref{tab:obs}).  The largest pixel scale available on NSFCam, $0\farcs3$ pixel$^{-1}$, was used to provide a field of view of $76{\arcsec}{\times}76{\arcsec}$.  This large field enabled the detection of ultracool companions to separations up to 1000~AU for our nearest targets.  Each target was observed using two to three sets of five dither positions to allow for sky background subtraction and to minimize the effects of sky variability and detector defects.  

We reduced the NSFCam data using IDL and IRAF routines.  The five dither positions were subtracted, shifted, and combined to create a background subtracted composite image.  Finally, if there were multiple sets of dithers, we added the composite images together to create a final image.  Candidates were limited to have separations from their primary between $2\arcsec$ and $35{\arcsec}-40{\arcsec}$ on average.  The outer limit is set by the edge of the image and the inner limit by the size of the PSF of the primary.  We identified candidates by eye and obtained relative photometry of each object in the field from the final composite images using the {\it qphot} script within IRAF.  The relative magnitudes of each source were estimated using published magnitudes of the target primaries (references are listed in Table \ref{tab:obs}).  

We inserted faint artificial point sources uniformly across each composite image to determine its sensitivity.  Each image was searched for these sources by eye.  We found that the sensitivity of the array is uniform from outside the PSF of the primary (${\sim}2\arcsec$) almost to the edge of the chip (${\sim}38{\arcsec}$).  However, we also discovered that our initial data reduction procedures introduced artifact sources into the outer $7\arcsec$ of each image.  As a result, we revised the outer limit of our survey inward to $31\arcsec$ and rejected any candidates with larger projected separations.  We therefore assign each final composite image a uniform detection limit out to $31\arcsec$.  

\begin{figure}
\centering
\rotatebox{-90}{
\epsscale{0.5}
\plotone{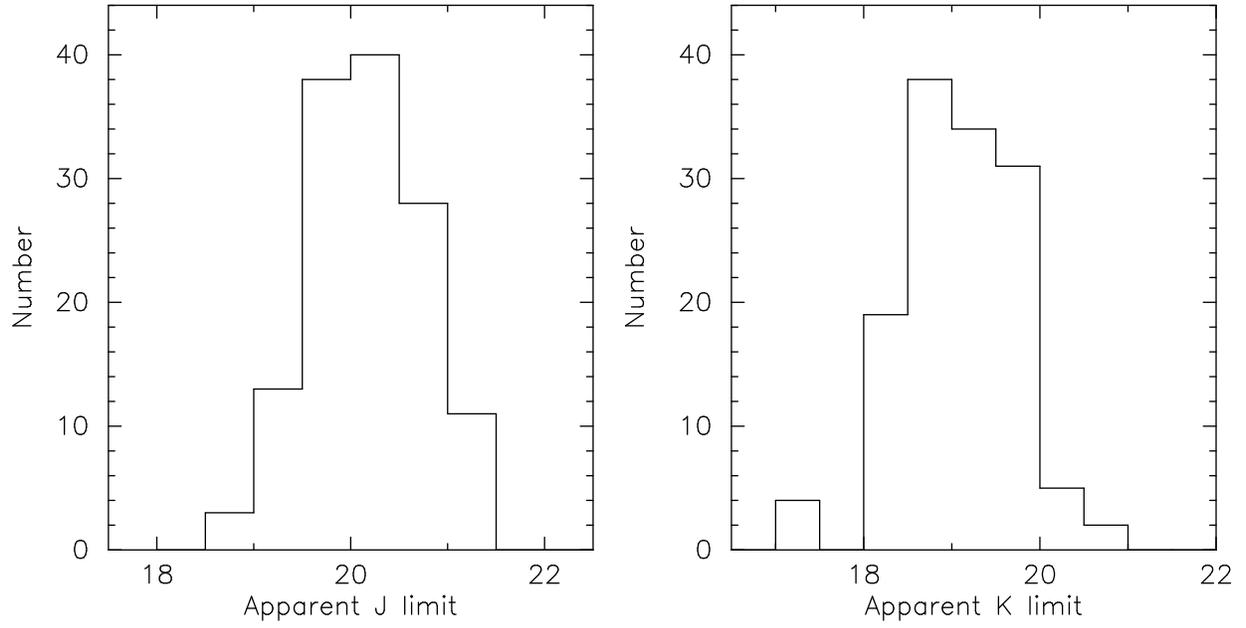}}
\caption{Histograms of the apparent $J$ (left) and $K$ (right) limiting magnitudes of the 132 first epoch NSFCam fields.}
\label{fig:ml}
\end{figure}

Figure \ref{fig:ml} shows the distribution of apparent magnitude limits in $J$ and $K$ for the survey fields; the median limiting magnitudes are $J=20.5$ and $K=18.5$.  Figure \ref{fig:dm} shows these limits expressed as companion detection limits (${\Delta}J$, ${\Delta}K$), the magnitude difference between the target and the detection limit.  Finally, Figure \ref{fig:al} transforms these sensitivity limits to the absolute magnitudes of potential secondary companions, where we show the location of Gl 229B as a reference.  Clearly, our observations extend well into the T dwarf regime and beyond in all cases.  In general, the sensitivity at $J$ is better than $K$ (particularly for neutral colored T dwarfs).  The $K$ band limits listed in Table \ref{tab:obs} therefore represent a conservative estimate of the sensitivity of our survey.

\subsection{WIYN Data}

Deep $I$ band images were obtained of a number of targets using the Mini-Mosaic Camera \citep{saha} on the WIYN telescope at Kitt Peak National Observatory.  The observations were made in August 2002 and February 2003.  Conditions were adequate, with seeing of 0.75 - 1 arcseconds.  While the August run was not photometric, the M and L dwarf targets provide an approximate local zero point that is sufficiently accurate to separate background objects from real companions, as discussed in the following section.  We used exposure times of 300 seconds, achieving typical limiting magnitudes for these observations of $I {\sim} 22.5-23$.

The images from the Mini-Mosaic Camera have a much larger field of view than the IRTF data.  We trimmed and rotated each frame to match the NSFCam field.  The relative photometry is based on the $I$ magnitudes estimated for the M and L dwarf primaries from $I-J$ colors given in Figure 4 of \citet{dahn}.  We estimated the $I$ magnitudes because few of our targets have published photometry in the $I$-band.  The relative photometry was measured in the same manner used for the NSFCam data ({\it qphot}).  

\section{Candidate Companion Selection}
\label{sec:selcrit}

\subsection{Near Infrared Criteria}

The candidate selection method was a multi-step process.  The initial step used the $M_{J}, J-K$ color-magnitude diagram.  Figure \ref{fig:sel} plots data for M, L, and T dwarfs with known trigonometric parallaxes.  We have used those objects to delineate the regions of the $M_{J}, J-K$ plane where we would expect to find low-luminosity companions to the ultracool targets.  L dwarfs have colors redder than $J-K$ = 1, and are brighter than $M_J \sim 15.5$; classical T dwarfs are bluer than $J-K$ = 0.5 and fainter than $M_J = 14$; and transitional, early-type T dwarfs have intermediate colors, and $14 < M_J < 15.5$.

We identify candidate companions by plotting color-magnitude data for each infrared source as if it were at the same distance as the appropriate ultracool target.  We use a $M_J$ versus spectral type relation to derive distances to all the target primaries that lack a trigonometric parallax.  This is the vast majority of our sample, only 6/132 have trigonometric parallax measurements.  The \citet{kc03} relation has distances uncertainties of ${\sim}10\%$, which corresponds to an uncertainty of ${\sim}\pm0.2$ mag.  If the source falls between the dashed and dotted lines plotted in Figure \ref{fig:sel}, then it is a potential low-luminosity companion.  A total of 221 sources meet these criteria.

\subsection{Optical Criteria}

Once the infrared candidates are selected, we cross-reference each against the POSS and UKST blue and red plates, as scanned in the Digital Sky Survey \citep{dposs}.  These photographic plates have limiting magnitudes of B$\sim$22 and R$\sim$21, while cool L and T dwarfs have extremely red optical-to-infrared colors, (R-J)$>$6 \citep{golim98}.  Thus, any sources visible on the DSS scans can be ruled out as candidate companions.  Thirty-six objects pass this criterion, see Figure \ref{fig:sel}, with colors consistent with late-L and T dwarfs.

Twenty-four of these remaining candidates were then observed at WIYN (Table \ref{tab:fu}).  With $I$-band observations of these objects, a new dimension is added to the color analysis.  In Figure 4 of \citet{dahn}, it is shown that dwarfs with spectral types of late-L and later have $I-J$ greater than 3.8.  Of the 24 objects observed at WIYN 23 were detected, 21 with $I-J$ colors less than 2.7 and 2 were discovered to be elongated (Table \ref{tab:fu}).  The remaining object, near 2M1146+22, was not detected and thus remained a viable candidate.  

An additional seven sources are within the field of the Sloan Digital Sky Survey, Fifth Data Release (DR5).  According to \citet{chiu}, L and T dwarfs have $i-z$ colors greater than 2 and $z-J$ colors greater than 2.5.  Six of the seven sources were detected in DR5 and all have $i-z$ and $z-J$ colors less than 0.8 and 1.7 respectively, and, therefore, are not ultracool companions.  This leaves 7 candidate companions.  Six of these remaining candidates have colors consistent with late-L or L/T transition dwarfs (Figure \ref{fig:sel}), and only one with T dwarf colors.  The T dwarf is rejected through methane band imaging, see Section \ref{sec:meth}.  Thus, our survey is a complete null result for T dwarf companions and nearly complete for L dwarfs.  We believe that the remaining objects are most likely not ultracool companions, but background stars given their position in the color-magnitude diagram.  Their $J-K$ colors correspond to main sequence K and M stars (see the right-hand panel of Figure 1 in \citet{kc03}).  Further observations of these final candidates will be obtained at a later date; however, the overall results of the present investigation are not affected significantly by the indeterminate properties of these objects.

\subsection{Methane Absorption Test}
\label{sec:meth}
The one remaining T dwarf candidate is a potential tertiary member of the 2M1146+22 system, which is a known, near equal-mass ultracool binary \citep{k99}.  The observed multi-epoch Keck fields are too small to cover the new candidate.  If a true brown dwarf companion, the candidate would be the faintest known brown dwarf, with an absolute $J$ magnitude of 18.4, approximately one magnitude fainter than the coolest known T dwarfs, such as Gl~570D \citep{geb01}.  It would also be the widest known brown dwarf multiple system, with a separation from the known binary of $21\arcsec$, or $\sim$570~AU at $\sim$27~pc.

The candidate was imaged in $H_{MK}$ band and in the narrow 1.7~$\mu$m methane band filter, Spencer 1.7, during an NSFCam run at IRTF in April 2004.  Figure \ref{fig:fil} shows a typical late-T dwarf spectrum with the $H_{MK}$ band and Spencer 1.7 filter profiles.  The center of the Spencer 1.7 filter is on the 1.7 $\mu$m methane feature that is prominent in cool brown dwarf spectra.  Thus, we expect that objects with significant methane absorption will show a drop in flux from $H_{MK}$ to Spencer 1.7.

Table \ref{tab:ratio} lists the expected $H_{MK}$ to Spencer 1.7 flux ratios, as derived for known L and T dwarfs.  The values have been computed from flux calibrated near-infrared spectra.  All T dwarf ratios were calculated from spectra downloaded from Adam Burgasser's T dwarf archive (http://web.mit.edu/ajb/www/tdwarf).  The L dwarf ratios were calculated from Ian McLean's BDSS archive \citep{mcl}.  The ratio values are flat for L dwarfs ($\sim$3.5) and increase from $\sim4$ for early-type T dwarfs to $\sim$11 for the T7 dwarf, 2M0348-60.  If the candidate is a late-type T dwarf, as indicated by its $J-K$ color, then its $H_{MK}$/Spencer 1.7 flux ratio should be on the order of 10 (Table \ref{tab:ratio}).  

To calculate the flux ratio of the candidate, the ratio is calibrated for the main binary system, the L3/L3 2M1146+22.  Since no flux standards were observed, we use flux calibrated spectra of objects similar to that of the primary.  The raw count rate of the primary and the candidate companion were measured at both $H_{MK}$ and Spencer 1.7 using the method described in Section \ref{sec:datared}.  The measured raw count rate ratio for 2M1146+22 is $2.2 \pm 0.1$ and the candidate companion is $1.7 \pm 1.1$.  The values derived from flux calibrated spectra of the L2 dwarf 2M0015+35 and the L4 dwarf Gliese 165B are 3.5 and 3.3 respectively \citep{mcl}.  These values are about 50\% higher than the raw ratio for 2M1146+22.  Hence, the ratio for the candidate is expected to be $\sim$50\% higher, raising it to $2.6 \pm 1.1$.  We surmise that it does not exhibit any significant methane absorption, as would be expected for a very late-type T dwarf.  

The observed ratio between $H_{MK}$ and Spencer 1.7 for the candidate is also similar to the ratio of the bandwidths of the two filters (${\Delta}H \sim 0.35~{\mu}m, {\Delta}S = {\sim}0.15~{\mu}m$), ${\sim}2.3$.  Since the InSb detectors have a relatively uniform response with wavelength, this is consistent with a flat spectrum source.  It is concluded that this candidate companion to 2M1146+22 is not a low-mass brown dwarf companion.

\section{Discussion}

We have completed a thorough, statistically well-defined search for wide companions to ultracool dwarfs.  Previous large-scale surveys \citep{bouy03,gizis03} used optical imaging and concentrated on searching for companions at small separations; our survey is the first to sample the full T dwarf regime at separations from a few tens to thousands of AU.  We can calculate an upper limit on the frequency of companions at those separations from our null results.  We use a basic Poisson distribution to determine the probability of getting a null detection given the number of observations: $Prob(Null) = \exp(-N_{obs}{\times}Freq)$, where $N_{obs}$ is the number of observations (132) and $Freq$ is the frequency of companions.  We determine a conservative upper limit when the probability of obtaining a null result falls below $5\%$.  This occurs at a companion frequency of $2.3\%$.  We will address the issue of overall ultracool dwarf companion frequency with greater detail in a forthcoming paper (Allen submitted).

\begin{figure}
\centering
\rotatebox{-90}{
\epsscale{0.5}
\plotone{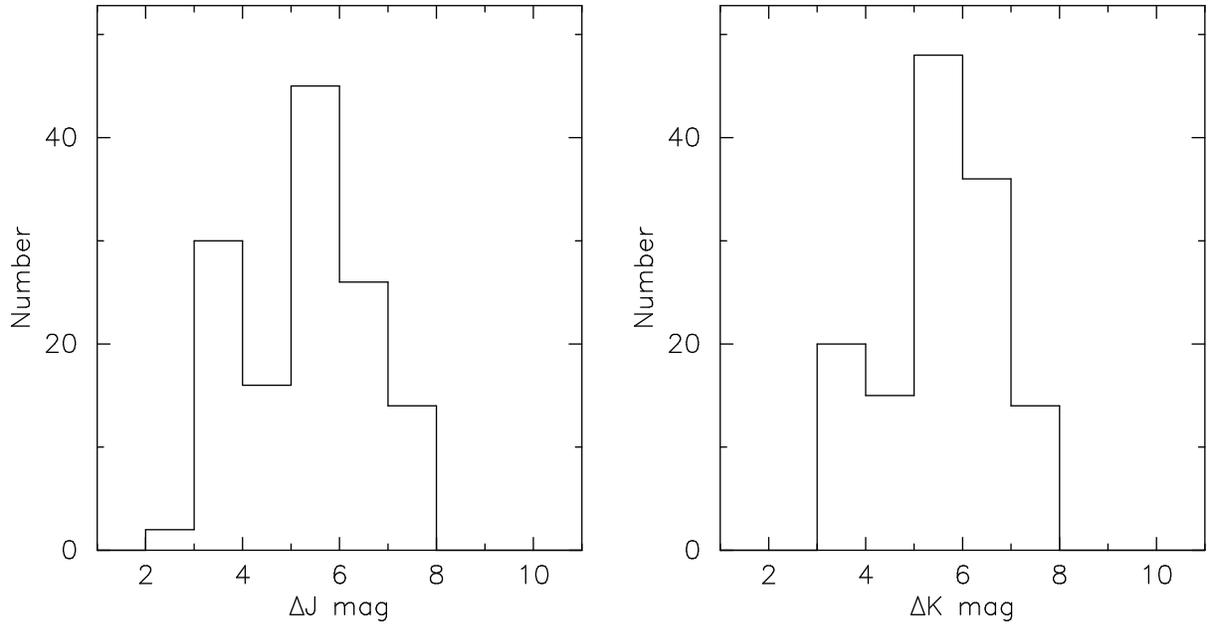}}
\caption{Histograms of the ${\Delta}J$ (left) and ${\Delta}K$ (right) companion detection limits of the 132 first epoch NSFCam fields.  We obtained the delta magnitudes by subtracting the magnitude of the target from the detection limit of each field.}
\label{fig:dm}
\end{figure}

\begin{figure}
\centering
\rotatebox{-90}{
\epsscale{0.5}
\plotone{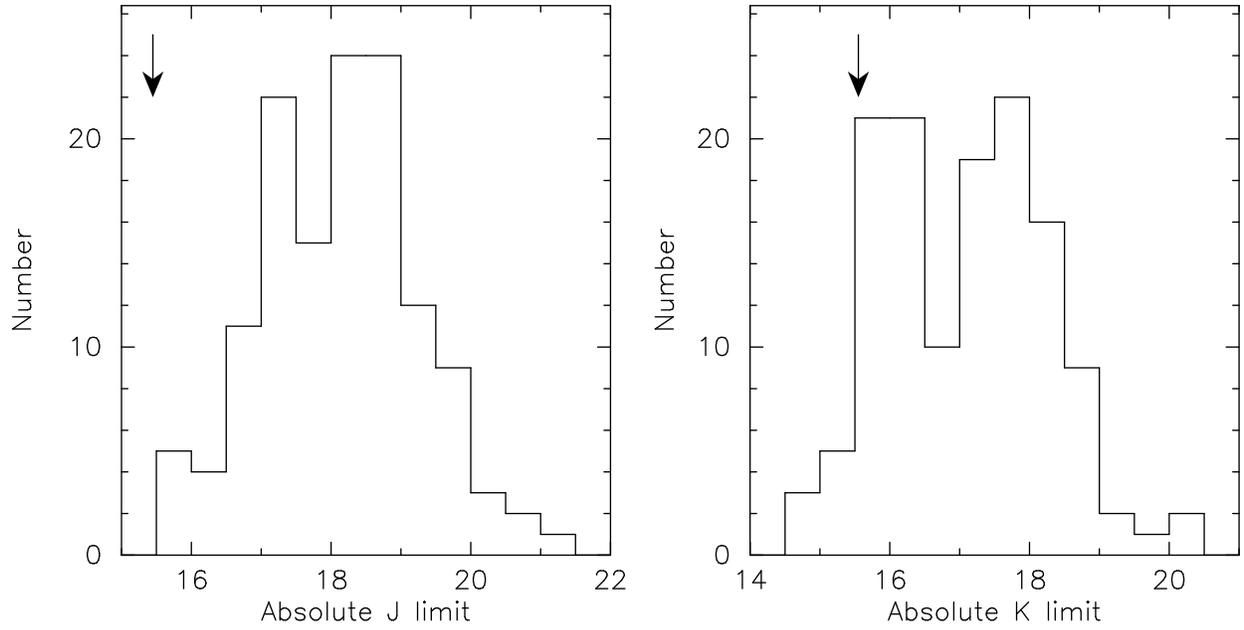}}
\caption{Histograms of the $J$ (left) and $K$ (right) limiting absolute magnitudes of the 132 first epoch NSFCam fields.  The position of the archetypal T dwarf Gl 229B is indicated with an arrow in both plots.}
\label{fig:al}
\end{figure}

\begin{figure}
\centering
\rotatebox{-90}{
\epsscale{0.8}
\plotone{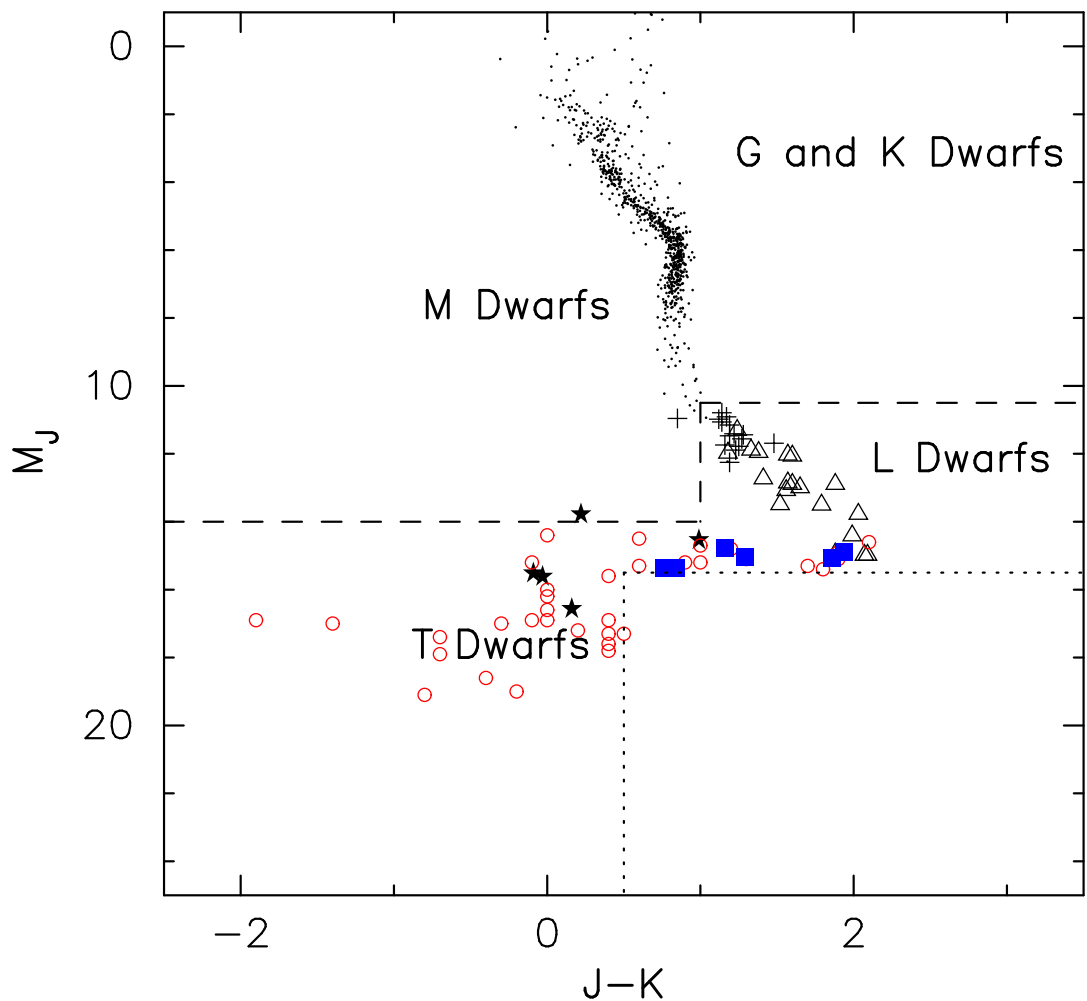}}
\caption{$M_J$ vs $J-K$ color magnitude diagram for nearby stars with trigonometric parallaxes: GKM dwarfs are shown as small points, ultracool M dwarfs as crosses, L dwarfs as open triangles, and T dwarfs as five-pointed stars.  The $M_J$ and $J-K$ selection criteria for candidate companions are shown as the dashed and dotted lines.  All sources that fall between those lines are accepted as initial candidates and are checked against the POSS plates.  The 36 objects that passed both criteria are plotted as open red circles at an absolute J magnitude that is consistent with the same distance as their putative primaries.  The six remaining candidates are marked as solid blue squares.}
\label{fig:sel}
\end{figure}

\begin{figure}
\centering
\rotatebox{-90}{
\epsscale{0.8}
\plotone{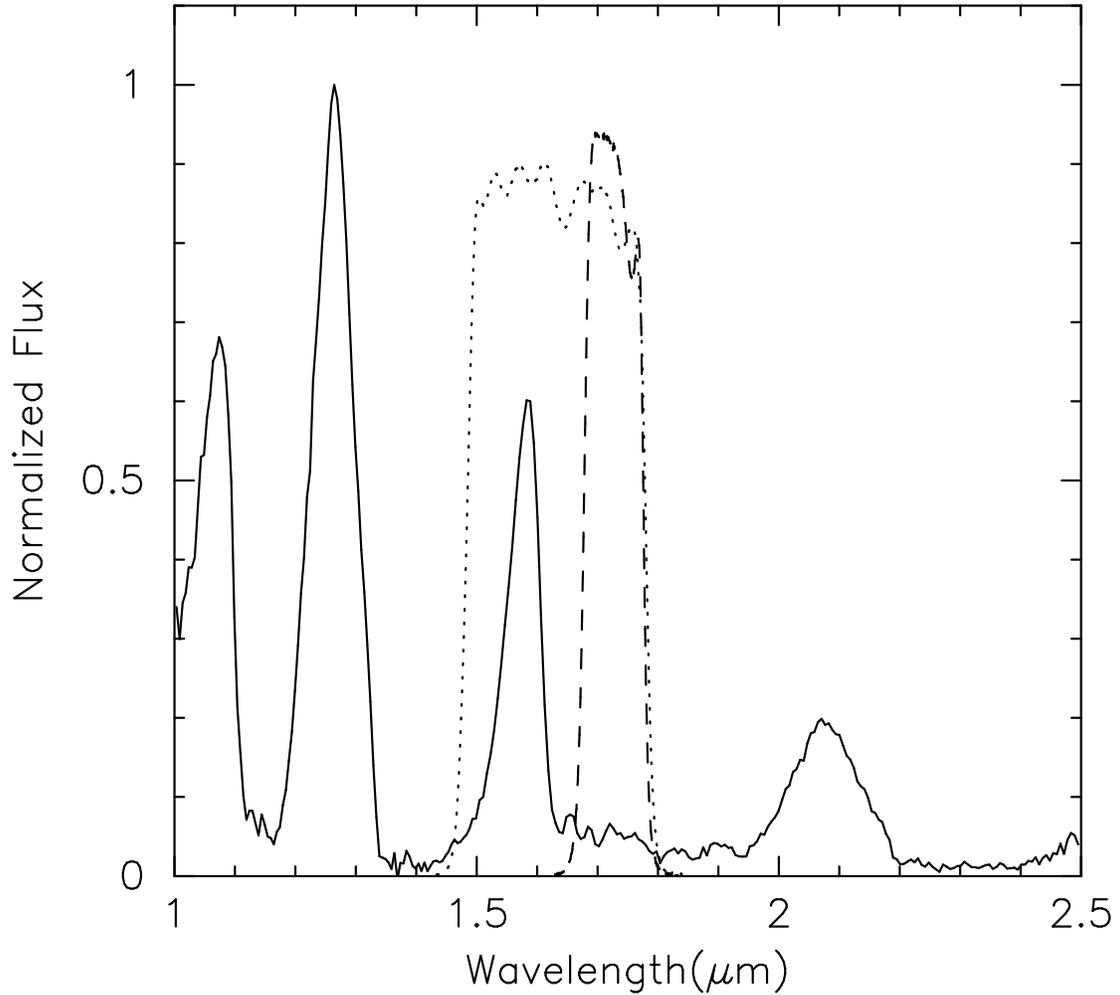}}
\caption{Near infrared spectrum of the T8 dwarf 2M0415-09 \citep{burg02} (solid) normalized such that the peak emission ($\sim$1.25~${\mu}$m) is equal to one, with the $H_{MK}$ (dotted) and Spencer 1.7 (dashed) filter transmission profiles.  The Spencer 1.7 filter falls in the 1.7 $\mu$m methane absorption feature making the $H_{MK}$ to Spencer 1.7 flux ratio indicative of methane absorption.}
\label{fig:fil}
\end{figure}

Since most of the targets of this survey are brown dwarfs, with masses that are dependent on the age of the system, we cannot directly express our results as an upper mass limit on wide companions.  However, as discussed in section 2.2, and illustrated in Figure \ref{fig:dm}, the average $K$-band limiting magnitude is $\sim$6 magnitudes fainter than the primary.  We can use the magnitude difference to obtain a statistical estimate of the likely limiting mass ratio for each system, $q = \frac{M_2}{M_1}$.

\begin{figure}
\centering
\rotatebox{-90}{
\epsscale{0.8}
\plotone{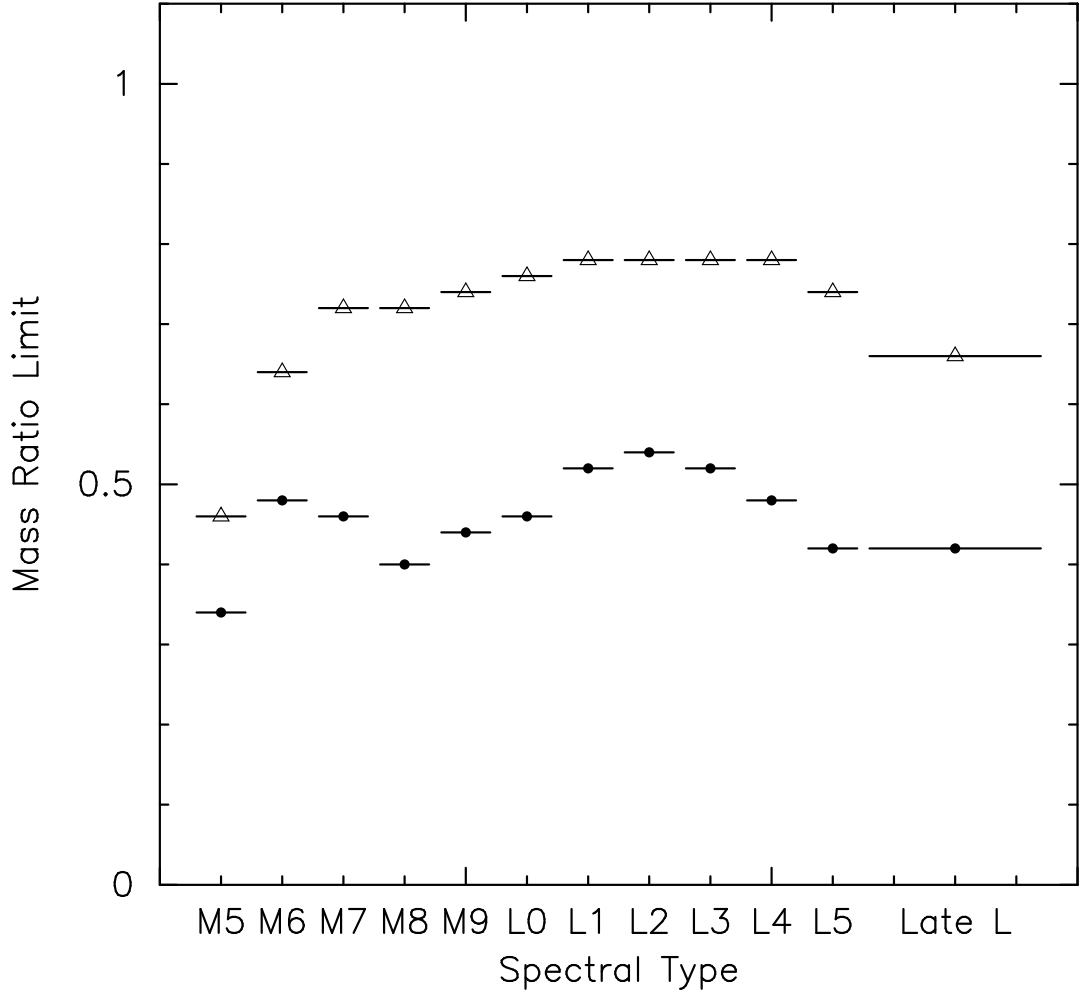}}
\caption{The predicted minimum mass ratio of detectable companions in the IRTF survey from our modeling as a function of primary spectral type for $85\%$ detection probability (triangles) and $50\%$ detection probability (circles).}
\label{fig:ql}
\end{figure}

We have transformed the observed magnitude difference to a mass ratio using the \citet{bur} low-mass star/brown dwarf evolutionary models.  The techniques used are similar to those employed in \citet{a05}; as in that paper, we transform the theoretical bolometric tracks to the M$_K$ plane using bolometric corrections from \citet{golim}.  Given M$_K$ and $\Delta$K for each source, we estimate the mass ratio detection limit for a range of ages from 20 Myr to 10 Gyr; the result is the detection probability of companions as a function of mass ratio and primary spectral type.  This detection probability is a measure of the likelihood that a companion of a given mass ratio can be detected.  An observation with a $60\%$ detection probability for a mass ratio of 0.4 means that we would find $60\%$ of the companions with a mass ratio of 0.4.  The detection probability increases with increasing mass ratio.  For example, 100$\%$ of companions with mass ratios between 0.8 and 1 are found in all observations.  Figure \ref{fig:ql} plots the mass ratio limit as a function of spectral type at which the detection probability falls below $85\%$ and $50\%$; the typical values are $q > 0.75$ and $q > 0.45$, respectively.  As discussed in \citet{a05}, M7 to L8 dwarfs are expected to have masses between 0.1 and 0.07 $M_{\odot}$.  Thus, these limits correspond to typical companion masses from 0.05 to 0.075~$M_{\odot}$ at $85\%$ detection probability and 0.03 to 0.05~$M_{\odot}$ at $50\%$.  Deeper imaging is required to probe the full range of potential mass ratios.  

The absence of wide companions to ultracool dwarfs has been discussed previously in the literature \citep{inr01,gizis03,bouy03,burg03}.  Our survey extends coverage to lower luminosities and lower mass ratio systems.  These results are generally consistent with the ejection scenario for brown dwarf formation \citep{rc01}, where only close, tightly-bound binary systems survive the ejection process.  However, recent hydrodynamic simulations by \citet{dcbh} suggest that dynamical disruption, rather than ejection, may be sufficient to account for the lack of wide, low-mass systems.  Moreover, \citet{burg03} have compiled data for a wide range of binary systems, and show that there is a correlation between maximum separations, $a_{max}$, and the total system mass, $M_{tot}$ (see their Figure 9) - at least for M$_{tot} < 1 M_{\odot}$.  Burgasser et~al also report a possible change in the boundary relation (defining $a_{max}$ as f($M_{tot}$)) from an exponential, $\log {a_{max}} \propto M_{tot}$, to $a_{max} \propto M_{tot}^2$ in the range $0.2 > M_{tot} > 0.1 M_\odot$.  This suggests that the absence of wide companions in very low-mass systems is the culmination of a continuous, mass-dependent mechanism, rather than a process specific to brown dwarf origins.

To summarize, we find that wide companions to ultracool dwarfs are rare, with a binary frequency upper limit of $2.3\%$, for companion masses above $0.03~M_{\odot} - 0.05~M_{\odot}$.  However, these results are one piece of the larger ultracool dwarf companion puzzle.  More extensive simulations and theoretical analyses, spanning the full mass range, are required to assess the full implications of the present results for brown dwarf formation scenarios.  This issue will be addressed in a future paper (Allen 2007, submitted), combining all extant ultracool dwarf companion surveys with observations of binary stars in the Solar Neighborhood.

\noindent {\it Acknowledgments}

P.R.A. acknowledges support by a grant made under the auspices of  the NASA/NSF NStars initiative, administered by JPL, Pasadena, CA.  P.R.A. also would like to thank Erika Nelson for her aid in the initial reduction of the IRTF NSFCam data.  The authors gratefully acknowledge and thank Abi Saha, Andrew Dolphin,  Rob Seaman, and Nelson Zarate for their help and support in acquiring the WIYN follow-up data.  We also would like to thank the support staff at the IRTF for their help in compiling the extensive series of observations discussed in this paper.

The Digitized Sky Survey was produced at the Space Telescope Science Institute under U.S. Government grant NAG W-2166. The images of these surveys are based on photographic data obtained using the Oschin Schmidt Telescope on Palomar Mountain and the UK Schmidt Telescope. The plates were processed into the present compressed digital form with the permission of these institutions.

The Second Palomar Observatory Sky Survey (POSS-II) was made by the California Institute of Technology with funds from the National Science Foundation, the National Aeronautics and Space Administration, the National Geographic Society, the Sloan Foundation, the Samuel Oschin Foundation, and the Eastman Kodak Corporation.

The UK Schmidt Telescope was operated by the Royal Observatory Edinburgh, with funding from the UK Science and Engineering Research Council (later the UK Particle Physics and Astronomy Research Council), until 1988 June, and thereafter by the Anglo-Australian Observatory. The blue plates of the southern Sky Atlas and its Equatorial Extension (together known as the SERC-J), as well as the Equatorial Red (ER), and the Second Epoch [red] Survey (SES) were all taken with the UK Schmidt.

\clearpage

\clearpage

\clearpage

\clearpage

\clearpage

\clearpage

\clearpage

\clearpage

\begin{deluxetable}{lcccccccc}
\tabletypesize{\footnotesize}
\rotate
\tablewidth{0pt}
\tablecaption{IRTF Observations}
\tablehead{
\colhead{Name} & \colhead{$J$} & \colhead{$K_s$} & \colhead{SpT} & \colhead{Distance(pc)} & \colhead{$m_{J_{lim}}$} & \colhead{$m_{K_{lim}}$} &\colhead{IRTF Obs Date} & \colhead{Ref.}
}
\startdata
2M0010$+$17 & 13.88$\pm$0.03 & 12.81$\pm$0.03 & M8   & 35.1     & 21.38 & 18.56 & Oct 2001		    &  1 \\
2M0015$+$35 & 13.82$\pm$0.04 & 12.81$\pm$0.03 & L2   & 20.1     & 19.57 & 20.31 & Aug 2000		    &  3 \\
2M0028$+$15 & 16.49$\pm$0.14 & 15.33$\pm$0.13 & L4.5 & 43.3     & 20.41 & 21.08 & Aug 2000		    &  3 \\
2M0030$-$14 & 16.79$\pm$0.16 & 15.36$\pm$0.09 & L7   & 29.7     & 20.71 & 19.72 & Aug 2000		    &  3 \\
2M0036$+$18 & 12.44$\pm$0.04 & 11.58$\pm$0.03 & L3.5 &{\it  8.8}& 20.69 & 19.83 & Aug 2000		    &  3 \\
2M0051$-$15 & 15.23$\pm$0.05 & 14.15$\pm$0.05 & L3.5 & 29.7     & 20.98 & 19.90 & Aug 2000		    &  3 \\
2M0058$-$06 & 14.32$\pm$0.03 & 13.45$\pm$0.04 & L0   & 33.0     & 20.07 & 19.20 & Aug 2000		    &  3 \\
2M0103$+$19 & 16.26$\pm$0.09 & 14.88$\pm$0.07 & L6   & 28.3     & 20.18 & 19.88 & Aug 2000		    &  3 \\
2M0104$+$14 & 13.70$\pm$0.02 & 12.66$\pm$0.03 & M8   & 32.2     & 21.23 & 19.93 & Aug 2000		    &  1 \\
2M0105$+$14 & 13.59$\pm$0.02 & 12.55$\pm$0.03 & M7   & 37.3     & 19.34 & 18.30 & Oct 2001		    &  1 \\
2M0109$+$29 & 12.92$\pm$0.02 & 11.70$\pm$0.02 & M9.5 & 18.4     & 21.17 & 19.20 & Oct 2001, Feb 2002	    &  1 \\
2M0130$+$17 & 13.66$\pm$0.03 & 12.58$\pm$0.02 & M8   & 31.7     & 19.41 & 20.08 & Oct 2001	      	    &  1 \\
2M0135$+$12 & 14.43$\pm$0.04 & 12.86$\pm$0.03 & L1.5 & 28.7     & 19.43 & 18.61 & Aug 2000, Feb 2002	    &  3 \\
2M0140$+$27 & 12.51$\pm$0.02 & 11.44$\pm$0.02 & M8.5 & 17.3     & 20.76 & 19.69 & Oct 2001, Feb 2002	    &  1 \\
2M0205$+$12 & 15.60$\pm$0.06 & 13.68$\pm$0.08 & L5   & 25.8     & 19.20 & 17.60 & Oct 2001	      	    &  3 \\
2M0208$+$25 & 16.21$\pm$0.03 & 14.41$\pm$0.04 & L1   & 69.6     & 20.57 & 18.77 & Aug 2000, Oct 2001, Feb 2002&3 \\
2M0208$+$27 & 15.70$\pm$0.07 & 13.87$\pm$0.06 & L5   & 27.0     & 20.70 & 19.62 & Oct 2001	      	    &  3 \\
2M0224$+$25 & 16.55$\pm$0.11 & 14.67$\pm$0.09 & L2   & 70.7     & 20.91 & 19.03 & Oct 2001	      	    &  3 \\
2M0240$+$28 & 12.62$\pm$0.02 & 11.62$\pm$0.02 & M7.5 & 21.4     & 20.12 & 19.12 & Oct 2001	            &  1 \\
2M0253$+$27 & 12.49$\pm$0.02 & 11.45$\pm$0.02 & M8   & 18.5     & 19.99 & 19.70 & Oct 2001, Feb 2002	    &  1 \\
2M0306$+$15 & 17.12$\pm$0.19 & 16.24$\pm$0.14 & L6   & 42.0     & 20.72 & 20.16 & Aug 2000, Feb 2002	    &  3 \\
2M0309$-$19 & 15.82$\pm$0.06 & 14.08$\pm$0.07 & L4.5 & 31.8     & 20.18 & 18.44 & Oct 2001	            &  3 \\
2M0326$+$29 & 15.23$\pm$0.06 & 13.62$\pm$0.06 & L3.5 &{\it 32.3}& 20.23 & 18.62 & Oct 2001	            &  2 \\
2M0328$+$23 & 16.67$\pm$0.14 & 15.62$\pm$0.13 & L8   & 24.3     & 20.27 & 19.22 & Aug 2000	            &  3 \\
2M0337$-$17 & 15.59$\pm$0.06 & 13.58$\pm$0.04 & L4.5 & 28.6     & 21.34 & 19.33 & Oct 2001		    &  3 \\
2M0350$+$18 & 12.95$\pm$0.02 & 11.76$\pm$0.02 & M9   & 19.8     & 20.45 & 19.26 & Oct 2001	            &  1 \\
2M0355$+$22 & 15.96$\pm$0.09 & 14.05$\pm$0.07 & L3   & 45.6     & 19.88 & 19.05 & Aug 2000, Feb 2002	    &  2 \\
2M0409$+$21 & 15.55$\pm$0.07 & 13.84$\pm$0.06 & L3   & 37.8     & 20.55 & 19.59 & Oct 2001, Feb 2002	    &  3 \\
2M0652$+$47 & 13.55$\pm$0.03 & 11.69$\pm$0.03 & L4.5 & 11.1     & 19.30 & 17.44 & Feb 2002	            &  4 \\
2M0708$+$29 & 16.75$\pm$0.12 & 14.69$\pm$0.09 & L5   & 43.8     & 20.67 & 19.05 & Oct 2001	            &  3 \\
2M0740$+$32 & 16.17$\pm$0.09 & 14.18$\pm$0.06 & L4.5 & 37.4     & 21.17 & 19.93 & Oct 2001	            &  3 \\
2M0746$+$20 & 11.74$\pm$0.03 & 10.49$\pm$0.03 & L0.5 &{\it 12.2}& 19.24 & 18.74 & Feb 2002	            &  3 \\
2M0753$+$29 & 15.49$\pm$0.05 & 13.85$\pm$0.06 & L2   & 43.4     & 19.41 & 18.85 & May 2001	            &  3 \\
2M0756$+$12 & 16.66$\pm$0.14 & 14.67$\pm$0.12 & L6   & 34.0     & 19.82 & 19.03 & May 2001, Feb 2002	    &  3 \\
2M0801$+$46 & 16.29$\pm$0.14 & 14.54$\pm$0.11 & L6.5 & 25.9     & 20.65 & 19.54 & Feb 2002	            &  3 \\
2M0810$+$14 & 12.71$\pm$0.02 & 11.61$\pm$0.02 & M9   & 17.8     & 20.21 & 19.11 & May 2001	            &  1 \\
2M0820$+$45 & 16.29$\pm$0.11 & 14.23$\pm$0.09 & L5   & 35.5     & 20.21 & 19.23 & May 2001, Feb 2002	    &  3 \\
2M0825$+$21 & 15.12$\pm$0.04 & 13.05$\pm$0.04 & L7.5 &{\it 10.7}& 20.87 & 18.80 & Feb 2002	            &  3 \\
2M0829$+$14 & 14.72$\pm$0.03 & 13.12$\pm$0.05 & L2   & 30.4     & 19.72 & 18.87 & May 2001		    &  3 \\
2M0829$+$26 & 17.08$\pm$0.20 & 14.81$\pm$0.10 & L6.5 & 37.3     & 20.08 & 19.17 & May 2001, Feb 2002	    &  3 \\
2M0832$-$01 & 14.13$\pm$0.03 & 12.69$\pm$0.03 & L1.5 & 25.0     & 19.88 & 18.44 & May 2001	            &  3 \\
2M0856$+$22 & 15.65$\pm$0.07 & 13.92$\pm$0.05 & L3   & 39.3     & 20.65 & 18.92 & Feb 2002		    &  4 \\
2M0914$+$22 & 15.06$\pm$0.04 & 13.68$\pm$0.03 & M9.5 & 49.2     & 20.81 & 19.43 & Feb 2002	            &  2 \\
2M0918$+$21 & 15.40$\pm$0.06 & 13.68$\pm$0.07 & L2.5 & 38.4     & 21.15 & 19.43 & Feb 2002	            &  2 \\
2M0925$+$17 & 12.60$\pm$0.02 & 11.60$\pm$0.02 & M7   & 23.6     & 20.85 & 19.10 & May 2001	            &  1 \\
2M0928$-$16 & 15.34$\pm$0.05 & 13.64$\pm$0.05 & L2   & 40.5     & 21.09 & 19.39 & May 2001, Feb 2002	    &  3 \\
2M0929$+$34 & 16.60$\pm$0.13 & 14.62$\pm$0.12 & L8   & 23.6     & 20.20 & 18.98 & May 2001, Feb 2002	    &  3 \\
2M0944$+$31 & 15.50$\pm$0.06 & 13.98$\pm$0.05 & L2   & 43.6     & 19.86 & 18.34 & May 2001	            &  3 \\
2M1017$+$13 & 14.10$\pm$0.03 & 12.71$\pm$0.03 & L2   & 22.9     & 19.85 & 18.46 & May 2001, Feb 2002	    &  4 \\
2M1029$+$16 & 14.31$\pm$0.04 & 12.61$\pm$0.04 & L2.5 & 23.3     & 20.06 & 20.11 & May 2001	            &  3 \\
2M1035$+$25 & 14.70$\pm$0.04 & 13.28$\pm$0.04 & L1   & 34.7     & 20.45 & 19.03 & May 2001	            &  3 \\
2M1102$-$23 & 17.04$\pm$0.19 & 14.79$\pm$0.09 & L4.5 & 55.8     & 20.64 & 19.15 & May 2001	            &  3 \\
2M1104$+$19 & 14.38$\pm$0.04 & 12.95$\pm$0.04 & L4   & 18.8     & 20.13 & 18.70 & May 2001	            &  4 \\
2M1112$+$35 & 14.57$\pm$0.04 & 12.69$\pm$0.05 & L4.5 &{\it 21.7}& 20.32 & 18.44 & Feb 2002	            &  3 \\
2M1123$+$41 & 16.07$\pm$0.08 & 14.34$\pm$0.06 & L2.5 & 52.3     & 20.43 & 20.09 & May 2001	            &  3 \\
2M1145$+$23 & 15.32$\pm$0.05 & 13.65$\pm$0.06 & L1.5 & 43.2     & 19.68 & 19.40 & Feb 2002	            &  2 \\
2M1146$+$22 & 14.03$\pm$0.03 & 12.44$\pm$0.03 & L3   &{\it 27.2}& 19.78 & 19.94 & Feb 2002	            &  2 \\
2M1155$+$23 & 15.72$\pm$0.07 & 14.12$\pm$0.06 & L4   & 33.7     & 20.72 & 19.12 & May 2001	            &  2 \\
2M1213$-$04 & 14.67$\pm$0.05 & 13.00$\pm$0.04 & L5   & 16.7     & 20.42 & 18.75 & Feb 2002	            &  4 \\
2M1218$-$05 & 14.06$\pm$0.03 & 12.74$\pm$0.03 & M8.5 & 35.3     & 19.81 & 18.49 & Feb 2002	            &  4 \\
2M1239$+$20 & 14.37$\pm$0.03 & 12.99$\pm$0.03 & M9   & 38.1     & 20.12 & 18.74 & Feb 2002	            &  2 \\
2M1239$+$55 & 14.67$\pm$0.03 & 12.74$\pm$0.03 & L5   & 16.8     & 20.42 & 20.24 & May 2001	            &  3 \\
2M1246$+$40 & 15.00$\pm$0.04 & 13.30$\pm$0.04 & L4   & 24.2     & 20.75 & 19.05 & May 2001	            &  3 \\
2M1254$+$25 & 14.36$\pm$0.07 & 13.24$\pm$0.09 & M7.5 & 47.8     & 20.11 & 18.99 & May 2001	            &  5 \\
2M1256$+$28 & 14.68$\pm$0.08 & 13.53$\pm$0.08 & M7.5 & 55.4     & 19.68 & 19.28 & May 2001, Feb 2002	    &  5 \\
2M1300$+$19 & 12.71$\pm$0.02 & 11.61$\pm$0.03 & L1   & 13.9     & 20.96 & 19.11 & May 2001	            &  1 \\
2M1328$+$21 & 16.00$\pm$0.11 & 13.99$\pm$0.09 & L5   & 32.3     & 20.36 & 19.74 & May 2001, Feb 2002	    &  2 \\
2M1332$+$26 & 16.11$\pm$0.10 & 14.36$\pm$0.08 & L2   & 57.7     & 21.11 & 19.36 & May 2001	            &  3 \\
2M1338$+$41 & 14.22$\pm$0.03 & 12.75$\pm$0.03 & L2.5 & 22.3     & 19.97 & 20.25 & May 2001	            &  3 \\
2M1343$+$39 & 16.18$\pm$0.08 & 14.11$\pm$0.06 & L5   & 33.7     & 21.18 & 19.11 & May 2001	            &  3 \\
2M1403$+$30 & 12.01$\pm$0.02 & 11.63$\pm$0.02 & M8.5 & 13.8     & 20.26 & 19.88 & May 2001	            &  1 \\
2M1411$+$39 & 14.68$\pm$0.04 & 13.27$\pm$0.05 & L1.5 & 32.2     & 20.43 & 20.77 & May 2001	            &  3 \\
2M1412$+$16 & 13.89$\pm$0.04 & 12.59$\pm$0.03 & L0.5 & 25.5     & 19.64 & 18.34 & May 2001	            &  3 \\
2M1421$+$18 & 13.21$\pm$0.02 & 11.93$\pm$0.02 & M9.5 & 21.0     & 20.71 & 20.18 & May 2001, Feb 2002	    &  1 \\
2M1426$+$15 & 12.87$\pm$0.02 & 11.71$\pm$0.02 & M9   & 19.1     & 21.12 & 19.96 & May 2001	            &  1 \\
2M1430$+$29 & 14.27$\pm$0.04 & 12.77$\pm$0.03 & L2   & 24.8     & 20.02 & 18.52 & May 2001	            &  4 \\
2M1438$+$64 & 12.99$\pm$0.02 & 11.65$\pm$0.02 & M9.5 & 18.4     & 21.24 & 19.15 & May 2001	            &  4 \\
2M1438$-$13 & 15.53$\pm$0.05 & 13.88$\pm$0.06 & L3   & 37.4     & 19.89 & 19.63 & May 2001	            &  3 \\
2M1439$+$18 & 16.12$\pm$0.11 & 14.73$\pm$0.11 & L1   & 66.8     & 20.04 & 19.09 & May 2001	            &  2 \\
2M1444$+$30 & 11.68$\pm$0.02 & 10.57$\pm$0.02 & M8   & 12.7     & 19.93 & 18.82 & May 2001	            &  1 \\
2M1449$+$23 & 15.80$\pm$0.08 & 14.34$\pm$0.10 & L0   & 65.2     & 20.80 & 20.09 & May 2001	            &  3 \\
2M1457$+$45 & 13.14$\pm$0.02 & 11.92$\pm$0.02 & M9   & 21.6     & 20.64 & 19.42 & May 2001	            &  1 \\
2M1506$+$13 & 13.41$\pm$0.03 & 11.75$\pm$0.03 & L3   & 14.1     & 19.16 & 19.25 & Aug 2000	            &  1 \\
2M1526$+$20 & 15.62$\pm$0.07 & 13.92$\pm$0.06 & L7   & 17.3     & 21.37 & 19.67 & May 2001	            &  3 \\
2M1546$+$37 & 12.44$\pm$0.02 & 11.42$\pm$0.02 & M7.5 & 19.7     & 20.69 & 19.67 & May 2001	            &  1 \\
2M1550$+$30 & 12.99$\pm$0.02 & 11.92$\pm$0.02 & M7.5 & 25.4     & 20.49 & 20.17 & May 2001	            &  1 \\
2M1551$+$64 & 12.87$\pm$0.02 & 11.73$\pm$0.02 & M8.5 & 20.4     & 21.12 & 19.23 & May 2001		    &  1 \\
2M1553$+$14 & 13.02$\pm$0.02 & 11.85$\pm$0.02 & M9   & 20.5     & 20.52 & 19.35 & Aug 2000, Oct 2001	    &  1 \\
2M1600$+$17 & 16.10$\pm$0.10 & 14.67$\pm$0.12 & L1.5 & 61.8     & 21.10 & 20.42 & May 2001	            &  3 \\
2M1615$+$35 & 14.55$\pm$0.04 & 12.89$\pm$0.05 & L3   & 23.8     & 20.30 & 18.64 & May 2001	            &  3 \\
2M1635$+$42 & 12.89$\pm$0.03 & 11.80$\pm$0.02 & M8   & 22.2     & 20.39 & 19.30 & May 2001	            &  1 \\
2M1656$+$28 & 17.10$\pm$0.20 & 14.96$\pm$0.16 & L4.5 & 57.3     & 21.46 & 19.32 & May 2001	            &  3 \\
2M1707$+$43 & 13.97$\pm$0.03 & 12.62$\pm$0.04 & L0.5 & 26.3     & 19.72 & 18.37 & Aug 2000, Oct 2001	    &  4 \\
2M1707$+$64 & 12.56$\pm$0.02 & 11.83$\pm$0.02 & M9   & 16.6     & 20.81 & 19.33 & May 2001	            &  1 \\
2M1710$+$21 & 15.74$\pm$0.08 & 14.19$\pm$0.09 & M8   & 82.6     & 20.10 & 18.55 & May 2001	            &  2 \\
2M1711$+$22 & 17.10$\pm$0.19 & 14.69$\pm$0.10 & L6.5 & 37.6     & 21.02 & 19.05 & May 2001, Feb 2002	    &  3 \\
2M1726$+$15 & 15.65$\pm$0.07 & 13.64$\pm$0.05 & L2   & 46.7     & 20.65 & 19.39 & May 2001, Feb 2002	    &  3 \\
2M1728$+$39 & 15.96$\pm$0.08 & 13.90$\pm$0.05 & L7   & 20.3     & 21.71 & 19.65 & May 2001, Feb 2002	    &  3 \\
2M1733$+$46 & 13.21$\pm$0.02 & 11.86$\pm$0.02 & M9.5 & 21.0     & 20.71 & 19.36 & Aug 2000, Oct 2001, Feb 2002&1 \\
2M1743$+$58 & 14.02$\pm$0.03 & 12.67$\pm$0.03 & M9.5 & 30.4     & 21.52 & 20.17 & May 2001, Feb 2002	    &  4 \\
2M1750$+$44 & 12.79$\pm$0.02 & 11.76$\pm$0.02 & M7.5 & 23.2     & 20.29 & 19.26 & Aug 2000	            &  1 \\
2M1841$+$31 & 16.12$\pm$0.10 & 14.97$\pm$0.08 & L4   & 40.5     & 20.48 & 19.97 & Aug 2000, Oct 2001	    &  3 \\
2M2049$-$19 & 12.87$\pm$0.02 & 11.77$\pm$0.02 & M7.5 & 24.1     & 20.37 & 19.27 & May 2001	            &  1 \\
2M2054$+$15 & 16.51$\pm$0.13 & 15.58$\pm$0.16 & L1   & 79.9     & 20.11 & 19.94 & Aug 2000, Oct 2001        &  3 \\
2M2057$+$17 & 16.11$\pm$0.11 & 15.21$\pm$0.13 & L1.5 & 62.1     & 19.71 & 19.57 & Aug 2000, Oct 2001	    &  3 \\
2M2140$+$16 & 12.94$\pm$0.03 & 11.78$\pm$0.03 & M8.5 & 21.1     & 20.44 & 17.53 & May 2001	            &  1 \\
2M2147$-$26 & 13.04$\pm$0.02 & 11.92$\pm$0.03 & M7.5 & 26.0     & 20.54 & 17.67 & Oct 2001	            &  1 \\
2M2147$+$14 & 13.84$\pm$0.03 & 12.65$\pm$0.03 & M8   & 34.4     & 21.34 & 20.15 & Aug 2000	            &  1 \\
2M2206$-$20 & 12.43$\pm$0.02 & 11.35$\pm$0.03 & M8   & 18.0     & 19.93 & 18.85 & Aug 2000	            &  1 \\
2M2208$+$29 & 15.82$\pm$0.09 & 14.09$\pm$0.08 & L2   & 50.5     & 20.82 & 19.84 & Aug 2000	            &  3 \\
2M2221$+$11 & 13.30$\pm$0.03 & 12.30$\pm$0.03 & M7.5 & 29.3     & 20.80 & 19.80 & Oct 2001	            &  1 \\
2M2224$-$01 & 14.05$\pm$0.03 & 12.80$\pm$0.03 & L4.5 & 11.4     & 19.80 & 20.30 & Aug 2000	            &  3 \\
2M2234$+$23 & 13.14$\pm$0.02 & 11.81$\pm$0.02 & M9.5 & 20.3     & 20.64 & 19.31 & Aug 2000	            &  1 \\
2M2244$+$20 & 16.53$\pm$0.13 & 13.97$\pm$0.07 & L6.5 & 28.9     & 20.89 & 19.72 & Oct 2001	            &  6 \\
2M2306$-$05 & 11.37$\pm$0.02 & 10.29$\pm$0.02 & M7.5 & 12.1     & 19.62 & 18.54 & Oct 2001	            &  1 \\
2M2331$-$04 & 12.94$\pm$0.02 & 11.93$\pm$0.03 & M8   & 22.8     & 21.19 & 20.18 & Oct 2001	            &  1 \\
2M2334$+$19 & 12.77$\pm$0.02 & 11.64$\pm$0.02 & M8   & 21.0     & 20.27 & 19.89 & Aug 2000	            &  1 \\
2M2347$+$27 & 13.19$\pm$0.02 & 12.00$\pm$0.02 & M9   & 22.1     & 20.69 & 19.50 & Aug 2000	            &  1 \\
2M2349$+$12 & 12.62$\pm$0.02 & 11.56$\pm$0.02 & M8   & 19.6     & 20.12 & 19.81 & Oct 2001	            &  1 \\
D0909$-$06  & 14.01$\pm$0.03 & 12.51$\pm$0.03 & L0   & 28.6     & 19.76 & 20.01 & May 2001	            &  7 \\
D1047$-$18  & 14.24$\pm$0.03 & 12.88$\pm$0.04 & L2.5 & 22.5     & 21.74 & 20.38 & May 2001	            &  8 \\
D1159$+$00  & 14.25$\pm$0.03 & 12.67$\pm$0.03 & L0   & 31.9     & 20.00 & 20.17 & May 2001	            &  8 \\
D1323$-$18  & 15.06$\pm$0.04 & 14.17$\pm$0.05 & L0   & 46.4     & 20.06 & 19.17 & Feb 2002	            &  8 \\
SD0330$-$00 & 15.29$\pm$0.05 & 13.83$\pm$0.05 & L2   & 39.6     & 21.04 & 19.58 & Oct 2001	            &  9 \\
SD0413$-$01 & 15.33$\pm$0.05 & 14.14$\pm$0.06 & L0   & 52.5     & 21.08 & 19.14 & Oct 2001	            &  9 \\
SD0539$-$00 & 13.99$\pm$0.03 & 12.53$\pm$0.03 & L5   & 12.3     & 21.49 & 20.03 & Oct 2001	            &  9 \\
SD1203$+$00 & 14.01$\pm$0.03 & 12.48$\pm$0.03 & L3   & 18.6     & 19.76 & 19.98 & May 2001	            &  9 \\
SD1326$-$00 & 16.11$\pm$0.07 & 14.23$\pm$0.07 & L8   & 18.8     & 20.03 & 19.23 & May 2001	            &  9 \\
SD1440$+$00 & 15.95$\pm$0.08 & 14.60$\pm$0.10 & L1   & 61.8     & 21.70 & 19.60 & May 2001	            &  9 \\
SD1515$-$00 & 14.18$\pm$0.03 & 13.14$\pm$0.03 & M7   & 48.9     & 19.93 & 18.89 & May 2001	            &  9 \\
SD1619$+$00 & 14.39$\pm$0.04 & 13.19$\pm$0.05 & L2   & 26.1     & 20.14 & 20.69 & May 2001	            & 10 \\
SD1636$-$00 & 14.59$\pm$0.04 & 13.41$\pm$0.04 & L0   & 37.4     & 20.34 & 19.16 & May 2001, Oct 2001	    &  9 \\
\enddata
\label{tab:obs}
\tablecomments{Distances in italics are derived from trigonometric parallaxes in \citet{dahn}.\\References: (1) \citet{newn}; (2) \citet{kp99}; (3) \citet{kirk00}; (4) \citet{kc03}; (5) \citet{kp97}; (6) \citet{dahn}; (7) \citet{xd99}; (8) \citet{em99}; (9) \citet{xf00}; (10) \citet{sh02}}

\end{deluxetable}

\begin{deluxetable}{lccccccc}
\tablewidth{0pt}
\rotate
\tablecaption{Color Selected Candidate Companions}
\tablehead{
\colhead{Name} & \colhead{SpT$_{pri}$} & \colhead{$M_{J_{sec}}$} & \colhead{$J-K_{sec}$} & \colhead{Distance(pc)} & \colhead{$\Delta$RA(\arcsec)} & \colhead{$\Delta$DEC(\arcsec)} & \colhead{Notes} }

\startdata

  2M0010$+$17 & M8   & 15.2 & -0.1 & 35.1 & -12.8 &  12.3 & WIYNa, I-J=2.2,  CUT\\
  2M0015$+$35 & L2   & 16.9 & -1.9 & 20.1 &  27.4 &  -9.3 & WIYNa, I-J=1.2,  CUT\\
  2M0028$+$15 & L4.5 & 16.0 &  0.0 & 43.3 &   9.3 & -14.1 & WIYNa, I-J=1.2,  CUT\\
  2M0109$+$29 & M9.5 & 16.2 &  0.0 & 18.4 & -25.7 &  31.2 & WIYNa, I-J=2.7,  CUT\\
  2M0140$+$27 & M8.5 & 17.3 &  0.4 & 17.3 & -14.2 &  28.4 & WIYNa, Elongated, CUT\\
  2M0208$+$25 & L1   & 14.8 &  1.2 & 69.6 &  28.2 & -14.6 & * \\
  2M0208$+$25 & L1   & 14.9 &  1.9 & 69.6 & -26.3 &  17.9 & * \\
  2M0224$+$25 & L2   & 15.4 &  0.8 & 70.7 &  30.9 &  27.7 & * \\
  2M0253$+$27 & M8   & 15.2 &  0.9 & 18.5 &   2.9 & -24.4 & WIYNa, I-J=1.7,  CUT\\
  2M0253$+$27 & M8   & 17.3 &  0.5 & 18.5 & -11.6 & -27.3 & WIYNa, I-J=1.5,  CUT\\
  2M0306$+$15 & L6   & 16.9 &  0.4 & 42.0 &  17.6 &   0.7 & WIYNf, I-J=1.5,  CUT\\
  2M0306$+$15 & L6   & 16.9 &  0.0 & 42.0 &  16.3 &   8.5 & WIYNf, I-J=1.5,  CUT\\
  2M0306$+$15 & L6   & 17.0 & -0.3 & 42.0 &  28.1 &   6.6 & WIYNf, I-J=1.4,  CUT\\
  2M0326$+$29 & L3.5 & 15.1 &  1.3 & 32.3 & -24.9 &  -1.5 & * \\
  2M0409$+$21 & L3   & 17.2 &  0.2 & 37.8 & -11.1 & -24.1 & WIYNa, Elongated, CUT\\
  2M0409$+$21 & L3   & 17.0 & -1.4 & 37.8 &   7.7 &  14.1 & WIYNa, I-J=1.0,  CUT\\
  2M0753$+$29 & L2   & 15.6 &  0.4 & 43.4 & -18.9 & -11.8 & WIYNf, I-J=1.2,  CUT\\
  2M0829$+$26 & L6.5 & 14.4 &  0.0 & 37.3 &  30.8 &  28.6 & SDSS, i-z=0.56, z-J=0.8, CUT\\
  2M0856$+$22 & L3   & 15.3 &  0.6 & 39.3 &   2.8 &  23.7 & SDSS, i-z=0.65, z-J=1.34, CUT\\
  2M0918$+$21 & L2.5 & 15.1 &  1.9 & 38.4 & -21.7 &  -7.3 & SDSS, Not Detected, *\\
  2M0918$+$21 & L2.5 & 15.4 &  1.8 & 38.4 &  -6.7 &  25.0 & SDSS, i-z=0.28, z-J=1.7, CUT\\
  2M1102$-$23 & L4.5 & 15.4 &  0.8 & 55.8 &   2.5 & -22.6 & *\\
  2M1146$+$22 & L3   & 17.8 &  0.4 & 27.2 &  21.1 &   0.6 & WIYNf, Not Detected\\
  2M1439$+$18 & L1   & 15.3 &  1.7 & 66.8 & -12.8 & -23.9 & SDSS, i-z=0.55 z-J=1.23 CUT\\
  2M1707$+$64 & M9   & 19.0 & -0.2 & 16.6 &  11.3 & -17.2 & WIYNa, I-J=-0.5, CUT\\
  2M1710$+$21 & M8   & 14.7 &  1.0 & 82.6 &  20.0 & -13.3 & SDSS, i-z=0.8 z-J=1.03 CUT\\
  2M1710$+$21 & M8   & 14.6 &  2.1 & 82.6 &  -7.6 &   1.1 & SDSS, i-z=0.7 z-J=1.26 CUT\\
  2M1711$+$22 & L6.5 & 17.4 & -0.7 & 37.6 &  29.2 &  22.0 & WIYNa, I-J=0.4,  CUT\\
  2M1728$+$39 & L7   & 17.6 &  0.4 & 20.3 &  -5.9 &  35.6 & WIYNa, I-J=2.2,  CUT\\
  2M1733$+$46 & M9.5 & 17.9 & -0.7 & 21.0 &  20.0 & -13.8 & WIYNa, I-J=1.4,  CUT\\
  2M1733$+$46 & M9.5 & 16.6 &  0.0 & 21.0 & -32.1 & -29.0 & WIYNa, I-J=-0.1, CUT\\
  2M1743$+$58 & M9.5 & 15.2 &  1.0 & 30.4 & -23.4 & -12.0 & WIYNa, I-J=1.5,  CUT\\
  2M2049$-$19 & M7.5 & 16.9 & -0.1 & 24.1 &   6.3 & -20.1 & WIYNa, I-J=1.9,  CUT\\
  2M2140$+$16 & M8.5 & 18.6 & -0.4 & 21.1 &  15.1 &  25.0 & WIYNa, I-J=1.0,  CUT\\
  2M2208$+$29 & L2   & 14.5 &  0.6 & 50.5 &  -1.8 & -22.0 & WIYNa, I-J=2.0,  CUT\\
  2M2224$-$01 & L4.5 & 19.1 & -0.8 & 11.4 & -13.9 &  18.2 & WIYNa, I-J=0.9,  CUT\\

\enddata
\label{tab:fu}

\tablecomments{SDSS: Object is in SDSS DR5 field.  WIYNa: Field observed with WIYN in August 2002.  WIYNf: Field observed with WIYN in February 2003.  All objects detected in either SDSS or with WIYN imaging have colors that are too blue to be consistent with an ultracool dwarf.  Objects with *'s in the notes column are candidates that still require follow-up observations to determine their nature.}

\end{deluxetable}

\clearpage

\begin{deluxetable}{ccc}
\tablewidth{0pt}
\tablecaption{$H_{MK}/{S17}$ Ratios for Known L and T Dwarfs}
\tablehead{
\colhead{Spectral Type} & \colhead{Object Name} & \colhead{$\frac{H_{MK}}{S17}$} }

\startdata
L2 & 2M0015+35     &  3.5 \\
L4 & Gliese 165B   &  3.3 \\
T0 & SDSS 0423-04  &  3.9 \\
T2 & SDSS 1254-01  &  4.2 \\
T5 & 2MASS 2254+31 &  5.8 \\
T6 & SDSS 1624+00  &  7.2 \\
T7 & 2MASS 0348-60 & 11.5 \\
T8 & Gliese 570D   &  9.6 \\
\enddata
\tablecomments{Ratios calculated from published flux calibrated spectra and filter profiles.}
\label{tab:ratio}

\end{deluxetable}

\end{document}